\documentclass[sigconf,natbib]{acmart}

\AtBeginDocument{%
  \providecommand\BibTeX{{%
    \normalfont B\kern-0.5em{\scshape i\kern-0.25em b}\kern-0.8em\TeX}}}

\setcopyright{acmcopyright}
\copyrightyear{2023}
\acmYear{2023}
\acmDOI{XX.XXXX/XXXXXXX.XXXXXXX}

\acmConference[XX]{XX}{XX}{XX}
\acmBooktitle{XX,X,X}
\acmPrice{15.00}
\acmISBN{XXX-X-XXXX-XXXX-X/23/08}

\usepackage{algorithm}
\usepackage{algorithmic}
\usepackage{multirow}
\usepackage{subfigure}
\usepackage{array}
\usepackage{hyperref}

\begin{document}

\title{Multi-Granularity Click Confidence Learning via Self-Distillation in Recommendation}



\author{Chong Liu, Xiaoyang Liu, Lixin Zhang, Feng Xia, Leyu Lin}
\email{nickcliu@tencent.com}
\affiliation{%
  \institution{Tencent}
  \country{China}
  }



\renewcommand{\shortauthors}{Chong Liu et al.}

\begin{abstract}
Recommendation systems rely on historical clicks to learn user interests and provide appropriate items. However, current studies tend to treat clicks equally, which may ignore the assorted intensities of user interests in different clicks. In this paper, we aim to achieve multi-granularity Click confidence Learning via Self-Distillation in recommendation (CLSD). Due to the lack of supervised signals in click confidence, we first apply self-supervised learning to obtain click confidence scores via a global self-distillation method. After that, we define a local confidence function to adapt confidence scores at the user group level, since the confidence distributions can be varied among user groups. With the combination of multi-granularity confidence learning, we can distinguish the quality of clicks and model user interests more accurately without involving extra data and model structures. The significant improvements over different backbones on industrial offline and online experiments in a real-world recommender system prove the effectiveness of our model. Recently, CLSD has been deployed on a large-scale recommender system, affecting over 400 million users.
\end{abstract}

\ccsdesc[500]{Information systems~Recommender systems}

\keywords{Recommender Systems, Self-Distillation, Click Confidence}


\maketitle

\section{Introduction}
\begin{figure}[th]
\centering
\includegraphics[width=0.5\textwidth]{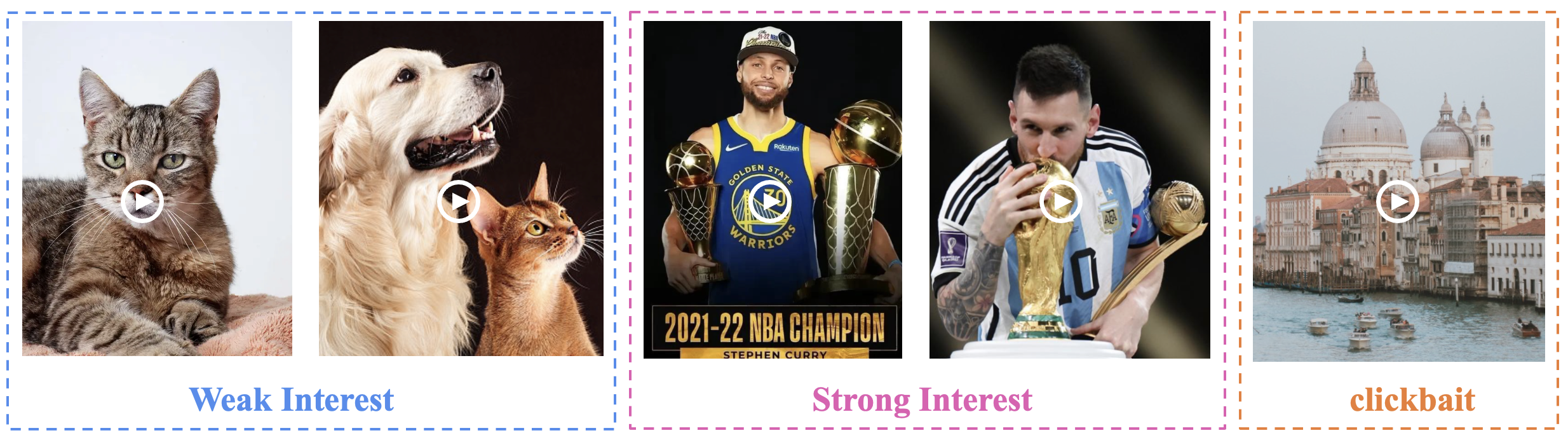}
\caption{An illustrative sample of historical clicks.}  \label{img:example}
\end{figure}

Online recommendation systems learn user interests and provide appropriate items based on historical interactions to maximize user satisfaction. Clicks are direct user feedbacks that can be easily collected from online serving.
Thus, the Click-Through Rate (CTR) prediction task is widely used in recommendation systems to model user interests \cite{LR,WideDeep,deepfm}.
Though promising, current works \cite{xdeepfm,autoint,dcn} that roughly treat all clicks equally may ignore the different intensities of user interests in clicks, since a click could be high-interested, low-interested, misleading-click or even clickbait for the user. 
Figure \ref{img:example} displays an example collected from the real world to reveal the different intensities of users' historical clicks. The user feedback indicates the inevitable disparity in click quality.

There are some works that focus on eliminating the effects of false-positive interactions\cite{wang2021denoising}, while we emphasize the existence of low-interested clicks beside noisy interactions. Some studies qualify clicks with additional user actions (i.e., favorite and skip) \cite{wen2019leveraging}, which may be sparse and hard to collect for most of the recommendation systems, especially for feed scenarios on which we mainly concentrate.
Generally, recommendation systems have the following two primary challenges to recognizing confidence levels of clicks:
1) The intensities of user interests in clicks are hard to qualify. Due to the lack of supervised signals, it is challenging to figure out high-quality clicks that really satisfy users from low-quality clicks, casual clicks, and clickbait. Though we can conduct customer surveys to obtain user preferences for clicks, the surveys of limited users are sparse and biased compared with our recommendation system's millions of users. 
2) The click confidence should be sample-level, which is varied from the related user and item. Current studies \cite{yi2014beyond} tend to apply dwell time to qualify click confidence and reflect user satisfaction. Since dwell time depends on many factors, i.e., the length of contents, user reading habits, and user interests, manually defining dwell time functions to qualify click confidence may involve extra data and noise.

In this paper, we aim to consider the influence of click confidence in CTR prediction without involving extra data or model structures. Concretely, we propose a simple yet effective method that achieves \textbf{multi-granularity Click confidence Learning via Self-Distillation in recommendation (CLSD)}. 
We obtain sample-level confidence scores from a dynamic teacher model via self-distillation.
Then, we conduct essential analyses of our confidence scores and personalize a local confidence function to adapt confidence scores at the user group level.
Finally, we jointly utilize the multi-granularity confidence learning to effectively train the model with varied emphases on different samples.
The advantages of CLSD are summarized as follows:
First, we apply self-supervised learning to distinguish the quality of clicks via self-distillation. 
Second, we generate sample-level confidence scores from the teacher model without involving extra data.
Finally, we can learn more accurate and reliable user interests by considering the distinction of confidence levels between samples.

We evaluate the performance of CLSD on real-world recommendation feed scenarios. CLSD achieves significant enhancements on both offline and online experiments, which verifies the effectiveness and universality of CLSD. The main contributions of our method are concluded as follows:
\begin{itemize}
  \item We emphasize the significance of click confidence, and propose our click confidence learning model. To the best of our knowledge, we are the first to utilize multi-granularity click confidence learning via self-distillation for CTR prediction.
  \item We explore a global distillation method and a local adaption module to achieve simple but efficient confidence learning. CLSD is a universal method with the model structure unchanged, which is easy to deploy in real-world recommendation systems.
  \item The significant improvements on both offline and online evaluations further prove the effectiveness of CLSD. Moreover, CLSD has been deployed on a real-world recommender system serving over 400 million users.
\end{itemize}

\section{Related Work}
\textbf{CTR Prediction.}
CTR prediction is increasingly essential to many real-world recommendation systems, and various methods \cite{nfm,afm,dcn,deepfm,xdeepfm,pnn,deepcross} have been proposed to enhance model performance. Among these methods, Logistic regression (LR) \cite{LR} and Factorization machine (FM) \cite{FM} focus on low-order feature interactions. They are both time-saving due to the linear model structure. Google develops a Wide\&Deep \cite{WideDeep} learning system, combining the advantages of both the linear shallow and deep models. After that, many deep learning techniques have been used in CTR prediction. The self-attention technique is the core of AutoInt \cite{autoint}, which can learn high-order feature interactions automatically. DFN \cite{dfn} utilizes a transformer to learn knowledge from both positive and negative feedback. Besides, some methods \cite{din,zhou2019deep,dsin,atrank,hpmn} concentrate on how to extract user preferences from historical sequence behaviors. 

\noindent\textbf{Self-Distillation.} 
Self-distillation research focuses on learning knowledge from the model itself \cite{furlanello2018born, chong2023ct4rec, liu2023ufnrec, liu2023future}.
Some studies learn knowledge in the aspect of input data, i.e., input features\cite{xu2019data} and class labels \cite{yun2020regularizing}.
\cite{zhang2019your} shrinks the size of the convolutional neural networks via self-distillation to enhance the performance.
\cite{kim2021self} progressively distills the model's knowledge to soften hard labels.
\cite{tarvainen2017mean} focuses on averaging model weights instead of predictions to achieve self-distillation.
Limited studies utilize self-distillation in the recommendation. 
\cite{huang2022two} introduces the knowledge distillation concept into GCN-based recommendation. 
In CLSD, we adopt self-distillation to confidence learning for CTR prediction.
\section{Model} \label{sec:method}
    
\subsection{Preliminary}
\begin{figure}[t]
    \centering
    \includegraphics[width=0.4\textwidth]{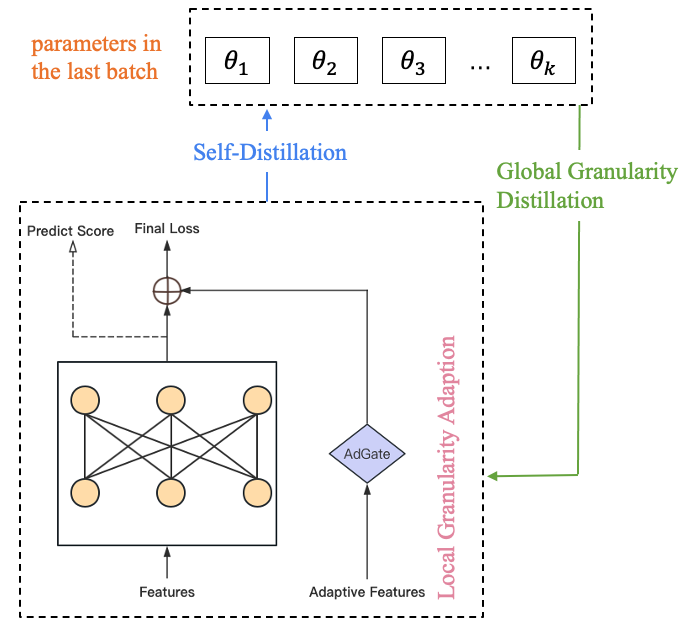}
    \caption{The model structure of CLSD.}  \label{img:main}
    \end{figure}

The overall structure of CLSD is presented as Figure~\ref{img:main}.
Before introducing our model, we introduce some basic notations in this part.  CTR prediction tasks aim to predict the next clicked items based on abundant sample features, which are mainly divided into multi-fields. For each sample pair $(u,i)$ in the training dataset \(\mathcal{D}\), \(\boldsymbol{x}\) denotes the corresponding features, and label \(y\in\left\{0,1\right\}\) denotes whether the user $u$ clicks the item $i$. 
The vector \(\boldsymbol{x}\) with multi-fields features can be denoted as \(\boldsymbol{x}=[\boldsymbol{x_{1}},\dots,\boldsymbol{x_{j}},\dots,\boldsymbol{x_{k}}]\),
where $\boldsymbol{x_{j}}$ represents the j-th field of \(\boldsymbol{x}\).
CTR prediction aims to model the probability \(p=f(x)\) that a user will click an item in a given context, where \(f(\cdot)\) indicates the model network. The universal loss function of CTR prediction is the negative log-likelihood loss to supervise model learning with labels:
\begin{equation}
\begin{aligned}
L_{\textit{ori}}=-\sum_{(u,i)} (y\log (p)+(1-y)\log (1-p)).
\end{aligned}
\label{eq.L_ori}
\end{equation}

\subsection{Global Granularity Distillation}
Due to the lack of supervised signals in click confidence, we introduce a self-supervised method to achieve confidence learning.
With the observation that noisy positive instances tend to have lower CTR prediction scores and bring larger loss than true positive instances~\cite{wang2021denoising}, we suppose that reliable CTR prediction scores of instances can reflect the user click confidence. Instead of directly using the model prediction, we involve a teacher model to provide more reliable and efficient scores.
Current self-supervised learning works provide numerous methods to obtain teacher models, i.e., a pre-trained large teacher model \cite{hinton2015distilling}, a previous model version with the best performance in the past epochs \cite{vu2021teaching}, or an exponential moving average of previous models \cite{tarvainen2017mean}.
However, these methods always need to store a set of additional model parameters in memory, which brings extra memory costs. To avoid the above problems and simplify the training process, we achieve self-distillation by directly choosing the model's current state to be its own teacher. 
Firstly, we warm up the model with the original training objective $L_{\textit{ori}}$.
After that, we regard the model generated from the last batch as the teacher for the current model, which can also alleviate the model forgetting issue~\cite{li2017learning}.
Then, the predictions of the teacher model are utilized to represent the confidence scores for the corresponding positive samples. Since we suppose the teacher model is a global-level user-interest learning ensemble, the predictions of the teacher model can reflect the intensities of user interests and confidence scores for different positive samples.
Thus, we scale the original loss with confidence scores for positive samples, which can be formulated as:
\begin{equation}
\begin{aligned}
L_{\textit{global}}=-\sum_{(u,i)} ((1+ p)y\log (p)+(1-y)\log (1-p)).
\end{aligned}
\label{eq.L_global}
\end{equation}
$p\in[0,1]$ is the prediction of the teacher model, serving as the confidence score. 

Our global granularity distillation step, short for Global-GD, has the following characteristics: 
1) Different from Focal loss \cite{lin2017focal} which down-weights the loss for well-classified samples, we up-weight the loss for well-classified samples and down-weight the loss for misclassified samples, since the click labels in recommendation systems are relatively uncertain with different confidence levels, while labels in CV are doubtless. Thus, we suppose users are actually more interested in their positive samples with higher predictions that are provided by a stable and well-trained teacher model.
2) Our confidence scores scale the original loss for only positive samples. In this paper, we do not focus on the intensities of dislikes for negative samples, so we only modify the original loss of positive samples. Besides, we maintain the scaled weights of positive samples not less than 1, due to the lack of positive samples compared with negative ones.
3) We dynamically adapt the original loss at the sample level upon confidence scores. Instead of bipolar values $\left\{0,1\right\}$, we treat the intensities of user interests in samples as continuous values in the range $[0,1]$, which is relatively reasonable and explainable. 
4) Compared with current studies \cite{yi2014beyond} using dwell time to qualify clicks, we utilize a self-distillation approach to avoid involving additional data which may lead to extra noise and the seesaw phenomenon.
Note that our Global-GD can match any self-distillation method. To save computation costs and memory costs, we achieve self-distillation via the current model method.

\subsection{Local Granularity Adaption}
The Global-GD mentioned above scales the original loss over the entire dataset from a general perspective. In this section, we will further analyze the impact of confidence learning at the user group level and define a local granularity adaption module.

\begin{figure}[th]
    \centering
    \includegraphics[width=0.5\textwidth]{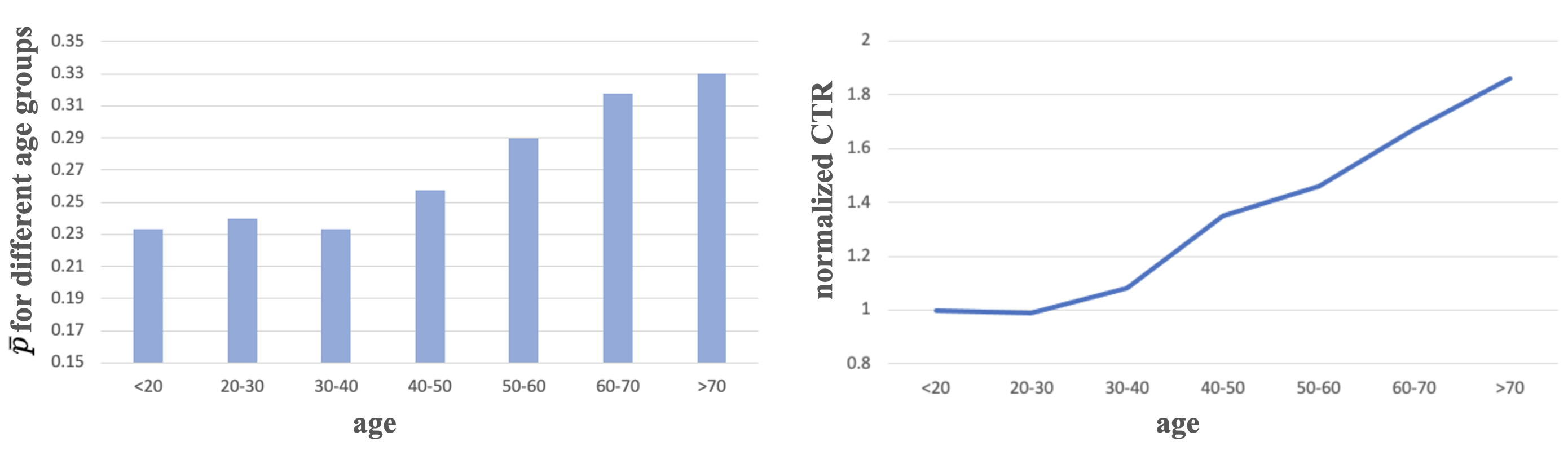}
    \caption{Distribution of $\bar{p}$ and CTR with age.}  \label{img:age_pre}
    \end{figure}

\noindent\textbf{Group Granularity Analysis.} 
Upon observing that confidence scores can vary among users, we analyze the prediction distribution of confidence scores $p$ at the user level and find that the different action patterns among user groups lead to a skewed distribution of $p$.
These user groups can be user profiles (e.g., gender, age), user interactive status (e.g., user activeness), or simply regard each user as a single group (e.g. individual level). In our real-world business scenarios, the impact of age groups on the final results is particularly significant. 
As shown in Figure~\ref{img:age_pre} (right), the normalized CTR collected from our offline dataset gradually increases with age. The trend of confidence scores $p$ in Figure~\ref{img:age_pre} (left) is similar to the normalized CTR. With our Global-GD module, the weights assigned to older users are higher due to their tendency to have higher $\bar{p}$, leading to an increasingly biased model towards this age group.
Therefore, we find that prior user information can influence the prediction distribution, involving the unfairness of Global-GD.

\noindent\textbf{Local Adaptive Module.}
In this part, we introduce an adaptive module to alleviate the bias caused by Global-GD. Concretely, we propose an Adaptive Gate, short for AdGate, to represent the imbalance of prior user information with personalized parameters and inject the adaptive results into the Global-GD. We denote the input features of the AdGate as adaptive features $x_{local}$, which can be user profiles ( i.e., age, location, gender and activeness). Since we observe the remarkable influence of the user's age on our industrial datasets, we directly take the age feature as $x_{local}$ in this paper to illustrate the effect of AdGate. Notice that the $x_{local}$ can be any prior features that bring inevitable bias to the distribution of the real CTR or the predicted CTR, as shown in Figure~\ref{img:age_pre}. Then, we formulate the output of AdGate as below:
\begin{equation}
\begin{aligned}
  p_{local} = \sigma(AdGate(x_{local}))
\end{aligned}
\label{eq.local}
\end{equation}
where AdGate is a simple Multi-Layer Perceptron (MLP) structure in this paper and can be directly extended to various complicated structures. The Sigmoid function, $\sigma(x)=1/(1+e^{-x})$, controls the output $p_{local}\in[0,1]$.
Here, we introduce $p_{local}$ to separately formulate the prediction of prior user information (i.e., age), which can alleviate the biased influence of $x_{local}$ on the training process of the backbone model as well as the Global-GD.

\subsection{Final Objective.} 
Finally, we combine the global granularity distillation and the local granularity adaption to obtain the final loss function:
\begin{equation}
\begin{aligned}
L_{\textit{final}}=-\sum_{(u,i)} ((1+\alpha p)y\log (p+p_{local})+(1-y)\log (1-p-p_{local})).
\end{aligned}
\label{eq.L_final}
\end{equation}
$\alpha\in[0,\infty]$ is a hyper-parameter to control the influence of Global-GD. Due to the relatively low ratio of positive samples among all samples, our weight of the positive samples is consistently not less than 1. 
To guarantee the performance of the teacher model, CLSD first warms up via $L_{\textit{ori}}$, empirically taking nearly $1/3$ of the entire training epochs, and then updates via $L_{\textit{final}}$.

\section{Experiments}
\subsection{Experiment Setup}
\begin{table}[th]
    \caption{Dataset statistics of two offline corpuses from real-world applications.}
    \centering
    \renewcommand\arraystretch{1.0}
    \begin{tabular}{p{2.5cm}<{\centering}|p{1.4cm}<{\centering}p{1.4cm}<{\centering}p{1.4cm}<{\centering}}
    \toprule
Dataset       & Instances & Fields & Features \\
\hline
Subscriptions & 762M      & 65     & 319M     \\
TopStory      & 614M      & 57     & 237M    \\
\bottomrule     
    \end{tabular}
    \label{tb:dataset}
\end{table}

\noindent\textbf{Datasets \& Settings.}
We conduct offline evaluations on two feed recommender systems, i.e., \emph{Subscriptions} and \emph{TopStory}. The detailed statistics of two offline datasets are shown in Table~\ref{tb:dataset}. 
For both datasets, we consider interactions in the first few days as train sets and the last day's interactions as test sets.
In experiments, all our models and baselines are optimized by Adam with the learning rate $0.003$. The batch sizes are $256$ for all models. 

\noindent\textbf{Baselines.}
We compared our method with eight models widely used in the industry:
(1) Wide\&Deep \cite{WideDeep}. With the development of deep models, Google improves greatly by combining a wide (or shallow) network and a deep one. Wide\&Deep is a general learning framework that can achieve the advantages of both wide and deep networks.
(2) DeepFM~\cite{deepfm}. DeepFM extends Wide\&Deep by substituting LR with FM to precisely model second-order feature interactions.
(3)DCN~\cite{dcn}. DCN introduces cross-network, which extends more feature crossing modes than FM. By controlling the number of layers, the cross-network can efficiently learn the low-dimensional feature interactions.
(4) AutoInt~\cite{autoint}. AutoInt automatically models the high-order interactions of input features by using self-attention networks.
(5) xDeepFM \cite{xdeepfm}. xDeepFM captures high-order interactions through its core module, the Compressed Interaction Network (CIN). CIN takes an outer product of a stacked feature matrix in a vector-wise way.
(6) DIEN \cite{zhou2019deep}. Dien introduces an interest evolving layer to capture the changing trend of the user interest. 
(7) CAN \cite{bian2022can}. CAN proposes a Co-Action Network (CAN) to effectively utilize the information of different feature pairs.


\subsection{Offline Evaluation}
\begin{table}[th]
\caption{Offline Evaluation on AUC (t-test with p$<$0.01).}
\begin{tabular}{p{1.35cm}<{\centering}|p{1.4cm}<{\centering}p{1.4cm}<{\centering}|p{1.4cm}<{\centering}p{1.4cm}<{\centering}}
\toprule
    & \multicolumn{2}{c|}{Subscriptions} & \multicolumn{2}{c}{TopStory} \\
Model   & Origin          & + CLSD          & Origin        & +CLSD        \\
\hline
W\&D    & 0.7662          & 0.7695          & 0.7874        & 0.7901       \\
DeepFM  & 0.7679          & 0.7708          & 0.7898        & 0.7929       \\
DCN     & 0.7681          & 0.7701          & 0.7892        & 0.7918       \\
AFM     & 0.7675          & 0.7709          & 0.7881        & 0.7906       \\
AutoInt & 0.7659          & 0.7702          & 0.7883        & 0.7907       \\
xDeepFM & 0.7687          & 0.7710          & 0.7904        & 0.7933       \\
DIEN    & 0.7680          & 0.7699          & 0.7884        & 0.7907       \\
CAN     & 0.7702          & 0.7725          & 0.7910        & 0.7940       \\
\hline
\end{tabular}
\label{tb:offline}
\end{table}


To verify the effectiveness and universality of CLSD, we apply CLSD to eight backbone models for offline evaluation. From Table \ref{tb:offline}, we can find that: 1) CAN performs better than other baselines, which indicates the Co-Action Network can effectively capture feature interaction for training. 
Moreover, our CLSD can continuously improve the performance of CAN, since we can distinguish the intensities of user interests in clicks.
2) CLSD can enhance the performance over eight backbone models, achieving 0.36\% (on average) improvements on \emph{Subscriptions} and 0.34\% improvements on \emph{TopStory} in terms of AUC. The significant improvements (t-test with $p<0.01$) prove the universality and efficiency of CLSD, which is complementary to various model structures and continuously enhances performance.

\begin{table}[th]
\caption{Online A/B Test (t-test with p$<$0.01).}
\begin{tabular}{p{4cm}<{\centering}|p{1.65cm}<{\centering} p{1.65cm}<{\centering}}
\toprule
Online Scenario             & ACN    & ADT    \\
\hline
Subscriptions Article    & +2.116\% & +1.403\% \\
Subscriptions Video      & +1.976\% & +1.984\% \\
Top Story Article & +1.158\% & +1.400\% \\
Top Story Video   & +1.444\% & +1.665\% \\
\bottomrule 
\end{tabular}
\label{tb:online1}
\end{table}

\begin{table}[]
\caption{Comparison of Different Objectives (with p$<$0.01).}
\begin{tabular}{p{3cm}<{\centering}|p{1.55cm}<{\centering} p{1.55cm}<{\centering}}
\toprule
Method            & ACN      & ADT      \\
\hline
DT-Reweight       & -1.806\% & +2.413\% \\
Focal Loss        & +0.130\% & -0.075\% \\
MTL               & +1.629\% & +1.871\% \\
CLSD        & +1.976\% & +1.984\% \\
DT-Reweight + CLSD & +1.249\% & +2.896\% \\
MTL + CLSD         & +2.717\% & +3.014\% \\
\bottomrule 
\end{tabular}
\label{tb:online2}
\end{table}
\subsection{Online Evaluation}
\noindent\textbf{General Performance.}
To verify the online performance of CLSD, we deploy CLSD on four online scenarios with differences in domains and data scales. Concretely, we conduct A/B tests on Subscriptions Article (with 43.8 million users for 5 days), Subscriptions Video (with 3.7 million users for 4 days), Top Story Article (with 4.3 million users for 8 days), Top Story Video (with 3 million users for 10 days). We mainly focus on two online metrics: average click number per capita (ACN) and average dwell time (ADT). 
From Table \ref{tb:online1}, we can observe that: 
1) CLSD improves both ACN and ADT on all four online scenarios. The relative improvements ranging from 1.4\% to 2.1\% are remarkable for such stable and mature online recommendation scenarios with sufficient features, enormous samples, and advanced models.
2) The concurrent improvements of ACN and ADT imply that CLSD can accurately and efficiently capture user interests, which further enhances user satisfaction in our recommender scenarios.

\noindent\textbf{Comparison of different objectives.}
We also compare the performance of CLSD with other objectives, i.e., DT-Reweight using a dwell time function to reweight the origin loss~\cite{xie2023reweighting}, focal loss \cite{lin2017focal}, and a multi-task learning loss with an additional dwell time loss based on Eq.\ref{eq.L_ori}. 
Online results on Subscriptions Video are presented in Table \ref{tb:online2}. We can observe that:
1) Compared with the above three objectives, CLSD achieves the best online performance in both ACN and ADT, which indicates CLSD can avoid the seesaw phenomenon in traditional reweighed objectives by focusing on click confidence learning. Since focal loss gives priority to hard samples, the focal loss may pay exceeding attention to weak interests or even casual clicks, leading to inferior performance.
2) With the combination of CLSD and the corresponding objectives, CLSD can continuously improve the performance of DT-reweight and MTL, which further proves the university and effectiveness of CLSD.

\subsection{Ablation Study}
\begin{table}[]
\caption{Ablation Study (t-test with p$<$0.01).}
\begin{tabular}{p{1.7cm}<{\centering}|p{1.35cm}<{\centering} p{1.35cm}<{\centering} p{1.35cm}<{\centering} p{1.35cm}<{\centering}}
\toprule
Model               & DeepFM & DCN    & AutoInt & CAN    \\
\hline
baseline            & 0.7679 & 0.7681 & 0.7659  & 0.7702 \\
+ local      & 0.7686 & 0.7689 & 0.7671  & 0.7711 \\
+ global            & 0.7697 & 0.7695 & 0.7686  & 0.7721 \\
+ CLSD              & 0.7708 & 0.7701 & 0.7702  & 0.7725 \\
\bottomrule
\end{tabular}
\label{tb:ablation}
\end{table}


As shown in Table \ref{tb:ablation}, we conduct an offline ablation study on four baselines on \emph{Subscriptions} to clarify the effect of different components of CLSD, i.e., the global granularity distillation and the local granularity adaption. Generally, the global component contributes more to the final performances, which implies the power of confidence learning via self-distillation. Also, the local component can adapt the loss at the user group level to further enhance the performance.

\subsection{Hyper-Parameter Analysis}
\begin{table}[th]
\caption{Analysis of $\alpha$ on Subscriptions.}
\begin{tabular}{l|llllll}
\toprule
$\alpha$            & 0      & 0.25   & 0.5    & 0.75   & 1.0    & 1.25   \\
\hline
CLSD & 0.7679 & 0.7690 & 0.7697 & 0.7703 & 0.7708 & 0.7692 \\
\bottomrule
\end{tabular}
\label{tb:alpha}
\end{table}
\noindent
In this section, we analyze the impact of the critical hyper-parameter $\alpha$ in Eq.~\ref{eq.L_final}, which controls the influence of CLSD during the training process.
Tabel~\ref{tb:alpha} shows the experiment results of CLSD with different $\alpha$ on \emph{Subscriptions}.
We can observe that a small $\alpha$ can bring a slight performance improvement on AUC. The performance improvement continuously increases with the increment of $\alpha$ and reaches the best at around $\alpha=1$. With a larger $\alpha$, the model excessively focuses on adapting the training process upon click confidence and ignores the original objective, leading to worse performance.

\section{Conclusion}
In this work, we propose a simple but effective multi-granularity click confidence learning method via self-distillation for CTR prediction. CLSD achieves remarkable improvements on both offline and online experiments with different backbones. Also, we analyze the distribution of our confidence scores for a deeper understanding. Besides, CLSD has been deployed on a real-world recommender system serving over 400 million users. In the future, we will explore more sophisticated local adaption methods in CLSD.




\bibliographystyle{ACM-Reference-Format}
\bibliography{ref}

\clearpage
\appendix







\end{document}